\documentstyle[a4,11pt]{article} 

\input amssym.def       
\input amssym.tex

\def\double{\Bbb}

\def\cc{{\double C}}     
       
\def\zz{{\double Z}}

\def\rr{{\double R}}

\newtheorem{theorem}{Theorem}

\newtheorem{definition}[theorem]{Definition}
\newtheorem{proposition}[theorem]{Proposition}

\def\si{\sigma}
\def\Si{\Sigma}
\def\cinf{C^{\infty}}
\def\cinfc{C^{\infty}_c}
\newcommand{\be}{\begin{equation}}
\newcommand{\ee}{\end{equation}}
\newcommand{\beq}{\begin{eqnarray}}
\newcommand{\eeq}{\end{eqnarray}}
\newcommand{\om}{\omega}
\newcommand{\Om}{\Omega}
\newcommand{\al}{\alpha}

\newcommand{\la}{\lambda}

\newcommand{\non}{\nonumber}

\newcommand{\ch}{\mbox{ch}}

\newcommand{\Ac}{{\cal A}}

\def\deb{\overline{\partial}}
\def\Id{\mbox{Id}}
\def\zb{\overline{z}}
\def\yb{\overline{y}}
\def\d{\partial}
\def\db{\overline{\partial}}
\def\Hc{{\cal H}}
\def\omb{\overline{\omega}}
\def\xb{\overline{x}}
\def\Bc{{\cal B}}

\begin{document}

\begin{center}

{\large A RIEMANN-ROCH THEOREM FOR ONE-DIMENSIONAL COMPLEX GROUPOIDS\\}
\vskip 1cm
{\bf Denis PERROT\footnotemark[1]}
\vskip 0.5cm
Centre de Physique Th\'eorique, CNRS-Luminy,\\ Case 907, 
F-13288 Marseille cedex 9, France \\[2mm]
{\tt perrot@cpt.univ-mrs.fr}\\[2mm]
\end{center}
\vskip 1cm
\begin{abstract} We consider a smooth groupoid of the form $\Si\rtimes\Gamma$ where $\Si$ is a Riemann surface and $\Gamma$ a discrete pseudogroup acting on $\Si$ by local conformal diffeomorphisms. After defining a $K$-cycle on the crossed product $C_0(\Si)\rtimes\Gamma$ generalising the classical Dolbeault complex, we compute its Chern character in cyclic cohomology, using the index theorem of Connes and Moscovici. This involves in particular a generalisation of the Euler class constructed from the modular automorphism group of the von Neumann algebra $L^{\infty}(\Si)\rtimes\Gamma$.
\end{abstract}

\vskip 1cm

\footnotetext[1]{Allocataire de recherche MENRT.}

\noindent {\bf I. Introduction}\\

In a series of papers \cite{CM95,CM98}, Connes and Moscovici proved a general index theorem for transversally (hypo)elliptic operators on foliations. After constructing $K$-cycles on the algebra crossed product $C_0(M)\rtimes\Gamma$, where $\Gamma$ is a discrete pseudogroup acting on the manifold $M$ by local diffeomorphisms \cite{CM95}, they developed a theory of characteristic classes for actions of Hopf algebras that generalise the usual Chern-Weil construction to the non-commutative case \cite{CM98,CM99}. The Chern character of the concerned $K$-cycles is then captured in the periodic cyclic cohomology of a particular Hopf algebra encoding the action of the diffeomorphisms on $M$. The nice thing is that this cyclic cohomology can be completely exhausted as Gelfand-Fuchs cohomology and renders the index computable.\\
We shall illustrate these methods with a specific example, namely the crossed product of a Riemann surface $\Si$ by a discrete pseudogroup $\Gamma$ of local conformal mappings. We find that the relevant characteristic classes are the fundamental class $[\Si]$ and a cyclic 2-cocycle on $\cinfc(\Si)\rtimes\Gamma$ generalising the (Poincar\'e dual of the) usual Euler class. When applied to the $K$-cycle represented by the Dolbeault operator of $\Si\rtimes\Gamma$, this yields a non-commutative version of the Riemann-Roch theorem. Throughout the text we also stress the crucial role played by the modular automorphism group of the von Neumann algebra $L^{\infty}(\Si)\rtimes\Gamma$. \\

\vskip 1cm
\newpage
\noindent {\bf II. The Dolbeault $K$-cycle}\\

Let $\Si$ be a Riemann surface without boundary and $\Gamma$ a pseudogroup of local conformal mappings of $\Si$ into itself. We want to define a $K$-cycle on the algebra $C_0(\Si)\rtimes\Gamma$ generalising the classical Dolbeault complex. Following \cite{CM95}, the first step consists in lifting the action of $\Gamma$ to the bundle $P$ over $\Si$, whose fiber at point $x$ is the set of K\"ahler metrics corresponding to the complex structure of $\Si$ at $x$. By the obvious correspondence metric $\leftrightarrow$ volume form, $P$ is the $\rr^*_+$-principal bundle of densities on $\Si$. The pseudogroup $\Gamma$ acts canonically on $P$ and we consider the crossed procuct $C_0(P)\rtimes\Gamma$.\\
Let $\nu$ be a smooth volume form on $\Si$. As in \cite{C1}, this gives a weight on the von Neumann algebra $L^{\infty}(\Si)\rtimes\Gamma$ together with a representative $\si$ of its modular automorphism group. Moreover $\si$ leaves $C_0(\Si)\rtimes\Gamma$ globally invariant and one has
\be
C_0(P)\rtimes\Gamma = (C_0(\Si)\rtimes\Gamma)\rtimes_{\si}\rr\ ,
\ee
where the space $P$ is identified with $\Si\times\rr$ thanks to the choice of the global section $\nu$. Therefore one has a Thom-Connes isomorphism \cite{C0}
\be
K_i(C_0(\Si)\rtimes\Gamma) \rightarrow K_{i+1}(C_0(P)\rtimes\Gamma)\ ,\quad i=0,1\ ,
\ee
and we shall obtain the desired $K$-homology class on $C_0(P)\rtimes\Gamma$. The reason for working on $P$ rather than $\Si$ is that $P$ carries quasi $\Gamma$-invariant metric structures, allowing the construction of $K$-cycles represented by differential hypoelliptic operators \cite{CM95}.\\

More precisely, consider the product $P\times\rr$, viewed as a bundle over $\Si$ with 2-dimensional fiber. The action of $\Gamma$ extends to $P\times \rr$ by making $\rr$ invariant. Up to another Thom isomorphism, the $K$-cycle may be defined on $C_0(P\times\rr)\rtimes\Gamma=(C_0(P)\rtimes\Gamma)\otimes C_0(\rr)$. By a choice of horizontal subspaces on the bundle $P\times\rr$, one can lift the Dolbeault operator $\deb$ of $\Si$. This yields the horizontal operator $Q_H=\deb+\deb^*$, where the adjoint $\deb^*$ is taken relative to the $L^2$-norm given by the canonical invariant measure on $P\times\rr$ (see \cite{CM95} for details). Finally, consider the signature operator of the fibers, $Q_V=d_Vd_V^*-d_V^*d_V$, where $d_V$ is the vertical differential. Then the sum $Q=Q_H+Q_V$ is a hypoelliptic operator representing our Dolbeault $K$-cycle.\\
This construction ensures that the principal symbol of $Q$ is completely canonical, because related only to the fibration of $P\times\rr$ over $\Si$, and hence is invariant under $\Gamma$. Another choice of horizontal subspaces does not change the leading term of the symbol of $Q$. This is basically the reason why $Q$ allows to construct a spectral triple (of even parity) for the algebra $\cinfc(P\times\rr)\rtimes\Gamma$. \\
If $\Gamma=\Id$, then $C_0(P\times\rr)\rtimes\Gamma=C_0(\Si)\otimes C_0(\rr^2)$ and the addition of $Q_V$ to $Q_H$ is nothing else but a Thom isomorphism in $K$-homology\be
K^*(C_0(\Si))\rightarrow K^*(C_0(P\times\rr))
\ee
sending the classical Dolbeault elliptic operator $\deb+\deb^*$ to $Q$.\\ 

Now we want to compute the Chern character of $Q$ in the periodic cyclic cohomology $H^*(\cinfc(P\times\rr)\rtimes\Gamma)$ using the index theorem of \cite{CM98}. We need first to construct an {\it odd} cycle by tensoring the Dolbeault complex with the spectral triple of the real line $(\cinfc(\rr),L^2(\rr),i{\partial\over{\partial x}})$. In this way we get a differential operator $Q'=Q+i{\partial\over{\partial x}}$ whose Chern character lives in the cyclic cohomology of $(\cinfc(P)\rtimes\Gamma)\otimes \cinfc(\rr^2)$. By Bott periodicity it is just the cup product
\be
\ch_*(Q') = \varphi \# [\rr^2]
\ee
of a cyclic cocycle $\varphi\in HC^*(\cinfc(P)\rtimes\Gamma)$ by the fundamental class of $\rr^2$. The main theorem of \cite{CM98} states that $\varphi$ can be computed from Gelfand-Fuchs cohomology, after transiting through the cyclic cohomology of a particular Hopf algebra. We perform the explicit computation in the remaining of the paper.\\

\vskip 1cm

\noindent {\bf III. The Hopf algebra and its cyclic cohomology}\\

First we reduce to the case of a flat Riemann surface, since for any groupoid $\Si\rtimes\Gamma$ one can find a flat surface $\Si'$ and a pseudogroup $\Gamma'$ acting by conformal transformations on $\Si'$ such that $C_0(\Si')\rtimes\Gamma'$ is Morita equivalent to $C_0(\Si)\rtimes\Gamma$ (see \cite{CM98} and section V below).\\

Let then $\Si$ be a flat Riemann surface and $(z,\zb)$ a complex coordinate system corresponding to the complex structure of $\Si$. Let $F$ be the $Gl(1,\cc)$-principal bundle over $\Si$ of frames corresponding to the conformal structure. $F$ is gifted with the coordinate system $(z,\zb,y,\yb)$, $y,\yb\in \cc{}^*$. A point of $F$ is the frame
\be
(y\partial_z,\yb\d_{\zb})\quad \mbox{at}\ (z,\zb)\ .
\ee
The action of a discrete pseudogroup $\Gamma$ of conformal transformations on $\Si$ can be lifted to an action on $F$ by pushforward on frames. More precisely, a holomorphic transformation $\psi\in \Gamma$ acts on the coordinates by
\beq
z&\rightarrow& \psi(z)\qquad \mbox{Dom} \psi\subset F\\
y&\rightarrow& \psi'(z)y\ ,\qquad \psi'(z)=\partial_z\psi(z)\ .
\eeq
Let $\cinfc(F)$ be the algebra of smooth complex-valued functions with compact support on $F$, and consider the crossed product $\Ac= \cinfc(F)\rtimes\Gamma$. $\Ac$ is the associative algebra linearly generated by elements of the form $fU^*_{\psi}$ with $\psi\in\Gamma$, $f\in\cinfc(F)$, $\mbox{supp} f\subset\mbox{Dom}\psi$. We adopt the notation $U_{\psi}\equiv U^*_{\psi^{-1}}$ for the inverse of $U^*_{\psi}$. The multiplication rule
\be
f_1U^*_{\psi_1}f_2U^*_{\psi_2}=f_1\,(f_2\circ\psi_1) U^*_{\psi_2\psi_1}
\ee
makes good sense thanks to the condition supp$f_i\subset \mbox{Dom}\psi_i$. We introduce now the differential operators
\be
X=y\d_z\qquad Y=y\d_y\qquad \overline{X}=\yb\d_{\zb}\qquad \overline{Y}=\yb\d_{\yb}
\ee
forming a basis of the set of smooth vector fields viewed as a module over $\cinf(F)$. These operators act on $\Ac$ in a natural way:
\be
X.(fU^*_{\psi})= (X.f)U^*_{\psi}\ ,\qquad Y.(fU^*_{\psi})=(Y.f)U^*_{\psi}
\ee
and similarly for $\overline{X},\overline{Y}$. Remark that the system $(z,\zb)$ determines a smooth volume form ${{dz\wedge d\zb}\over{2i}}$ on $\Si$. This in turn gives a representative $\si$ of the modular automorphism group of $L^{\infty}(\Si)\rtimes\Gamma$, whose action on $\cinfc(\Si)\rtimes\Gamma$reads (cf. \cite{C2} chap. III)
\be
\si_t(fU^*_{\psi})=|\psi'|^{2it}fU^*_{\psi}\ ,\quad t\in\rr\ .
\ee
We let $D$ be the derivation corresponding to the infinitesimal action of $\si$:
\be
D=-i{d\over{dt}}\si_t|_{t=0}\qquad D(fU^*_{\psi})=\ln |\psi'|^2fU^*_{\psi}\ .
\ee
The operators $\delta_n,\overline{\delta}_n$, $n\ge 1$ are defined recursively
\be
\delta_n=\underbrace{[X,...[X}_{n},D]...]\qquad \overline{\delta}_n=\underbrace{[\overline{X},...[\overline{X}}_{n},D]...]\ . \label{del}
\ee
Their action on $\Ac$ are explicitly given by
\be
\delta_n(fU^*_{\psi})=y^n\d^n_z(\ln\psi')fU^*_{\psi}\ ,\qquad \overline{\delta}_n(fU^*_{\psi})=y^n\d^n_z(\ln\overline{\psi'})fU^*_{\psi}\ .
\ee
Thus $\delta_n,\overline{\delta}_n$ represent in some sense the Taylor expansion of $D$. All these operators fulfill the commutation relations
\beq
{[Y,X]}&=& X\qquad [Y,\delta_n]\ =\ n\delta_n\non\\
{[X,\delta_n]}&=& \delta_{n+1} \qquad [\delta_n,\delta_m]\ =\ 0
\eeq
and similarly for the conjugates $\overline{X},\overline{Y},\overline{\delta}_n$. Thus $\{X,Y,\delta_n,\overline{X},\overline{Y},\overline{\delta}_n\}_{n\geq 1}$ form a basis of a (complex) Lie algebra. Let $\Hc$ be its enveloping algebra. The remarkable fact is that $\Hc$ is a Hopf algebra. First, the coproduct $\Delta: \Hc\rightarrow \Hc\otimes\Hc$ is determined by the action of $\Hc$ on $\Ac$:
\be
\Delta h(a_1\otimes a_2)=h(a_1a_2)\quad \forall h\in\Hc, a_i\in\Ac\ .
\ee
One has 
\beq
\Delta X&=& 1\otimes X+X\otimes 1 +\delta_1\otimes Y\\
\Delta Y&=&1\otimes Y+Y\otimes 1 \qquad \Delta\delta_1\ =\ 1\otimes\delta_1 +\delta_1\otimes 1\ .\non
\eeq
$\Delta\delta_n$ for $n>1$ is obtained recursively from (\ref{del}) using the fact that $\Delta$ is an algebra homomorphism, $\Delta(h_1h_2)=\Delta h_1\Delta h_2$. Similarly for the conjugate elements.\\
The counit $\varepsilon :\Hc\rightarrow \cc$ satisfies simply $\varepsilon(1)=1$, $\varepsilon(h)=0$ $\forall h\neq 1$.\\
Finally, $\Hc$ has an antipode $S:\Hc\rightarrow\Hc$, determined uniquely by the condition $m\circ S\otimes\Id\circ \Delta=m\circ \Id\otimes S\circ\Delta =\eta\varepsilon$, where $m:\Hc\otimes\Hc\rightarrow\Hc$ is the multiplication and $\eta:\cc\rightarrow\Hc$ the unit of $\Hc$. One finds
\be
S(X)=-X+\delta_1 Y\quad S(Y)=-Y\quad S(\delta_1)=-\delta_1\ .
\ee
Since $S$ is an antiautomorphism: $S(h_1h_2)=S(h_2)S(h_1)$, the values of $S(\delta_n)$, $n>1$ follow.\\

We are interested now in the cyclic cohomology of $\Hc$ \cite{CM98,CM99}. As a space, the cochain complex $C^*(\Hc)$ is the tensor algebra over $\Hc$:
\be
C^*(\Hc)= \bigoplus_{n=0}^{\infty}\Hc^{\otimes n}\ .
\ee
The crucial step is the construction of a characteristic map
\be
\gamma: \Hc^{\otimes n}\rightarrow C^n(\Ac,\Ac^*)
\ee
from the cochain complex of $\Hc$ to the Hochschild complex of $\Ac$ with coefficients in $\Ac^*$ \cite{C2}. First $F$ has a canonical $\Gamma$-invariant measure $dv=dzd\zb{{dyd\yb}\over{(y\yb)^2}}$. This yields a trace $\tau$ on $\Ac$:
\beq
\tau(f)&=&\int_F f\, dv\qquad f\in\cinfc(F)\ ,\non\\
\tau(fU^*_{\psi})&=& 0\qquad \mbox{if}\ \psi\neq 1\ .
\eeq
Then the characteristic map sends the $n$-cochain $h_1\otimes ...\otimes h_n\in\Hc^{\otimes n}$ to the Hochschild cochain $\gamma(h_1\otimes...\otimes h_n)\in C^n(\Ac,\Ac^*)$ given by
\be
\gamma(h_1\otimes...\otimes h_n)(a_0,...,a_n)=\tau(a_0h_1(a_1)...h_n(a_n))\ ,\qquad a_i\in\Ac\ .
\ee
The cyclic cohomology of $\Hc$ is defined such that $\gamma$ is a morphism of cyclic complexes. One introduces the face operators $\delta^i:\Hc^{\otimes (n-1)}\rightarrow\Hc^{\otimes n}$ for $0\leq i\leq n$:
\beq
\delta^0(h_1\otimes...\otimes h_{n-1})&=& 1\otimes h_1\otimes...\otimes h_{n-1}\non\\
\delta^i(h_1\otimes...\otimes h_{n-1})&=& h_1\otimes...\otimes\Delta h_i\otimes...\otimes h_{n-1}\qquad 1\leq i\leq n-1\non\\
\delta^n(h_1\otimes...\otimes h_{n-1})&=& h_1\otimes...\otimes h_{n-1}\otimes 1
\eeq
as well as the degeneracy operators $\si_i:\Hc^{\otimes(n+1)}\rightarrow\Hc^{\otimes n}$
\be
\si_i(h_1\otimes...\otimes h_{n+1})= h_1\otimes...\varepsilon(h_{i+1})...\otimes h_{n+1}\qquad 0\leq i\leq n\ .
\ee
Next, the cyclic structure is provided by the antipode $S$ and the multiplication of $\Hc$. Consider the twisted antipode $\tilde{S}=(\delta\otimes S)\circ\Delta$, where $\delta:\Hc\rightarrow\cc$ is a character such that
\be
\tau(h(a)b)=\tau(a\tilde{S}(h)(b))\qquad\forall a,b\in\Ac\ .
\ee
This last formula plays the role of ordinary integration by parts. One finds:
\beq
\delta(1)&=&1\ ,\qquad \delta(Y)\ =\ \delta(\overline{Y})\ =\ 1\non\\
\delta(X)&=&\delta(\overline{X})\ =\ \delta(\delta_n)\ =\ \delta(\overline{\delta}_n)\ =\ 0\quad \forall n\geq 1\ .
\eeq
The definition implies $\tilde{S}^2=1$. Connes and Moscovici proved in \cite{CM99} that the latter identity is sufficient to ensure the existence of a cyclicity operator $\tau_n:\Hc^{\otimes n}\rightarrow\Hc^{\otimes n}$
\be
\tau_n(h_1\otimes...\otimes h_n)=(\Delta^{n-1}\tilde{S}(h_1))\cdot h_2\otimes...\otimes h_n\otimes 1\ ,
\ee
with $(\tau_n)^{n+1}=1$. Now $C^*(\Hc)$ endowed with $\delta^i,\si_i,\tau_n$ defines a cyclic complex. The Hochschild coboundary operator $b:\Hc^{\otimes n}\rightarrow\Hc^{\otimes(n+1)}$ is
\be
b=\sum_{i=0}^{n+1} (-)^i\delta^i
\ee
and Connes' operator $B:\Hc^{\otimes (n+1)}\rightarrow\Hc^{\otimes n}$ is
\be
B=\sum_{i=0}^n(-)^{ni}(\tau_n)^iB_0\qquad B_0=\si_n\tau_{n+1}+(-)^n\si_n\ .
\ee
They fulfill the usual relations $B^2=b^2=bB+Bb=0$, so that $C^*(\Hc,b,B)$ is a bicomplex. We define the cyclic cohomology $HC^*(\Hc)$ as the $b$-cohomology of the subcomplex of cyclic cochains. The corresponding {\it periodic} cyclic cohomology $H^*(\Hc)$ is isomorphic to the cohomology of the bicomplex $C^*(\Hc,b,B)$ \cite{C2}. Furthermore, the definitions of $\delta^i,\si_i,\tau_n$ imply that $\gamma$ is a morphism of cyclic complexes. Consequently, $\gamma$ passes to cyclic cohomology
\be
\gamma: HC^*(\Hc)\rightarrow HC^*(\Ac)\ ,
\ee
as well as to periodic cyclic cohomology
\be
\gamma: H^*(\Hc)\rightarrow H^*(\Ac)\ .
\ee
In fact we are not interested in the frame bundle $F$ but rather in the bundle of metrics $P=F/SO(2)$, where $SO(2)\subset Gl(1,\cc)$ is the group of rotations of frames. $P$ is gifted with the coordinate chart $(z,\zb,r)$ where the radial coordinate $r$ is obtained from the decomposition 
\be
y=e^{-r+i\theta}\qquad r\in\rr\ , \theta\in[0,2\pi)\ .
\ee
The pseudogroup $\Gamma$ still acts on $P$ by
\beq
z&\rightarrow& \psi(z)\qquad \zb\rightarrow\overline{\psi(z)}\non\\
r&\rightarrow& r-{1\over 2}\ln|\psi'(z)|^2\ .
\eeq
Define $\Ac_1=\Ac^{SO(2)}\subset\Ac$ the subalgebra of elements of $\Ac$ invariant under the (right) action of $SO(2)$ on $F$. $\Ac_1$ is canonically isomorphic to the crossed product $\cinfc(P)\rtimes\Gamma$. $P$ carries a $\Gamma$-invariant measure $dv_1=e^{2r}dzd\zb dr$, so that there is a trace on $\Ac_1$, namely
\beq
\tau_1(f)&=&\int_Pf\, dv_1\qquad f\in\cinfc(P)\non\\
\tau_1(fU^*_{\psi})&=&0\qquad \mbox{if}\ \psi\neq 1\ .
\eeq
Thus passing to $SO(2)$-invariants yields an induced characteristic map
\be
\gamma_1: HC^*(\Hc,SO(2))\rightarrow HC^*(\Ac_1) \label{bug}
\ee
from the relative cyclic cohomology of $\Hc$, with $\gamma_1(h_1\otimes...\otimes h_n)(a_0,...,a_n)=\tau_1(a_0h_1(a_1)...h_1(a_n))$, $a_i\in\Ac_1$, where $h_1\otimes...\otimes h_n$ represents an element of $HC^*(\Hc,SO(2))$. The map $\gamma_1$ generalises the classical Chern-Weil construction of characteristic classes from connexions and curvatures. In the crossed product case $\Si\rtimes\Gamma$, these classes are captured by the periodic cyclic cohomology of $\Hc$. The authors of \cite{CM98} computed the latter as Gelfand-Fuchs cohomology. This is the subject of the next section.\\
\vskip 1cm

\noindent {\bf IV. Gelfand-Fuchs cohomology}\\

Let $G$ be the group of complex analytic transformations of $\cc$. $G$ has a unique decomposition $G=G_1G_2$, where $G_1$ is the group of affine transformations
\be
x\rightarrow ax+b\ ,\qquad x\in\cc\ ,\ a,b\in\cc
\ee
and $G_2$ is the group of transformations of the form
\be
x\rightarrow x+o(x)\ .
\ee
Any element of $G$ is then the composition $k\circ\psi$ for $k\in G_1$, $\psi\in G_2$. Since $G_2$ is the left quotient of $G$ by $G_1$, $G_1$ acts on $G_2$ from the right: for $k\in G_1$, $\psi\in G_2$, one has $\psi\triangleleft k\in G_2$. Similarly, $G_2$ acts on $G_1$ from the left: $\psi\triangleright k\in G_1$.\\

Remark that $G_1$ is the crossed product $\cc\rtimes Gl(1,\cc)$. The space $\cc\times Gl(1,\cc)$ is a prototype for the frame bundle $F$ of a flat Riemann surface. This motivates the notation $a=y$, $b=z$ for the coordinates on $G_1$. Under this identification, the left action of $G_2$ on $G_1$ corresponds to the action of $G_2$ on $F$: for a holomorphic transformation $\psi\in G_2$, one has
\be
z\rightarrow \psi(z)\ ,\qquad y\rightarrow\psi'(z)y\ ,
\ee
with $\psi(0)=0$, $\psi'(0)=1$. Furthermore, the vector fields $X,\overline{X},Y,\overline{Y}$ form a basis of invariant vector fields for the left action of $G_1$ on itself, i.e. a basis of the (complexified) Lie algebra of $G_1$. Its dual basis is given by the left-invariant 1-forms (Maurer-Cartan form)
\beq
\om_{-1}&=& y^{-1}dz\qquad \omb_{-1}\ =\ \yb^{-1}d\zb\non\\
\om_0&=& y^{-1}dy\qquad \omb_0\ =\ \yb^{-1}d\yb\ .
\eeq
The left action $G_2\triangleright G_1$ implies a right action of $G_2$ on forms by pullback. One has in particular, for $\psi\in G_2$,
\be
\om_{-1}\circ\psi = \om_{-1}\qquad \om_0\circ\psi= \om_0+y\d_z\ln\psi'\om_{-1}\qquad \mbox{and c.c.}\label{pull}
\ee

Consider now the discrete crossed product $\Hc_*=\cinfc(G_1)\rtimes G_2$ where $G_2$ acts on $\cinfc(G_1)$ by pullback. As a coalgebra, $\Hc$ is dual to the algebra $\Hc_*$. One has a natural action of $\Hc$ on $\Hc_*$:
\beq
X.(fU^*_{\psi}) &=& X.fU^*_{\psi}\qquad f\in\cinfc(G_1),\psi\in G_2\ ,\non\\
\delta_n(fU^*_{\psi})&=& y^n\d^n_z\ln\psi'\, fU^*_{\psi}\ ,
\eeq
and so on with $Y,\overline{X}$... The operators $\delta_n,\overline{\delta_n}$ have in fact an interpretation in terms of coordinates on the group $G_2$: for $\psi\in G_2$, $\delta_n(\psi)$ is by definition the value of the function $\delta_n(U^*_{\psi})U_{\psi}$ at $1\in G_1$. For any $k\in G_1$, one has
\be
{[\delta_n(U^*_{\psi})U_{\psi}]}(k) = \delta_n(\psi\triangleleft k)\ .
\ee
Note that (\ref{pull}) rewrites
\be
\om_0\circ\psi = \om_0+\delta_1(\psi\triangleleft k)\om_{-1}\qquad \mbox{at}\ k\in G_1\ .
\ee
The Hopf subalgebra of $\Hc$ generated by $\delta_n,\overline{\delta_n}$, $n\geq 1$, corresponds to the commutative Hopf algebra of functions on $G_2$ which are {\it polynomial} in these coordinates.\\

Let $A$ be the complexification of the formal Lie algebra of $G$. It coincides with the jets of holomorphic and antiholomorphic vector fields of any order on $\cc$:
\beq
\d_x&,& x\d_x\ ,...,\ x^n\d_x\ ,...\qquad x\in\cc\non\\
\d_{\xb}&,& \xb\d_{\xb}\ ,...,\ \xb^n\d_{\xb}\ ,... 
\eeq
The Lie bracket between the elements of the above basis is thus
\beq
{[x^n\d_x,x^m\d_x]}&=& (m-n)x^{n+m-1}\d_x\qquad \mbox{and c.c.}\non\\
{[x^n\d_x,\xb^m\d_{\xb}]}&=& 0\ .
\eeq
Define the generator of dilatations $H=x\d_x+\xb\d_{\xb}$ and of rotations $J=x\d_x-\xb\d_{\xb}$. They fulfill the properties
\beq
{[H,x^n\d_x]}&=& (n-1)x^n\d_x\qquad [H,\xb^n\d_{\xb}]\ =\ (n-1)\xb^n\d_{\xb}\non\\
{[J,x^n\d_x]}&=& (n-1)x^n\d_x\qquad [J,\xb^n\d_{\xb}]\ =\ -(n-1)\xb^n\d_{\xb}\ .
\eeq
We are interested in the Lie algebra cohomology of $A$ (see \cite{G}). The complex $C^*(A)$ of cochains is the exterior algebra generated by the dual basis $\{\om^n,\omb^n\}_{n\geq -1}$:
\beq
\om^n(x^m\d_x)&=& \delta^m_{n+1}\qquad \om^n(\xb^m\d_{\xb})\ =\ 0\\
\omb^n(x^m\d_x)&=& 0\qquad \omb^n(\xb^m\d_{\xb})\ =\ \delta^m_{n+1}\qquad \forall n\geq -1,m\geq 0\ ,\non
\eeq
and the coboundary operator is uniquely defined by its action on 1-cochains
\be
d\om(X,Y)= -\om([X,Y])\qquad \forall X,Y\in A\ .
\ee
>From \cite{CM98} we know that the {\it periodic} cyclic cohomology $H^*(\Hc,SO(2))$ is isomorphic to the relative Lie algebra cohomology $H^*(A,SO(2))$, i.e. the cohomology of the basic subcomplex of cochains on $A$ relative to the Cartan operation $(L,i)$ of $J$:
\be
L_J\om = (i_Jd+di_J)\om\qquad \forall \om\in C^*(A)\ .
\ee
We say that a cochain $\om\in C^*(A)$ is of weight $r$ if $L_H\om=-r\om$. Remark that 
\be
L_H\om^n = -n\om^n\ ,\qquad L_H\omb^n=-n\omb^n\qquad  \forall n\geq -1\ ,
\ee
so that $C^*(A)$ is the direct sum, for $r\geq -2$, of the spaces $C^*_r(A)$ of weight $r$. Since $[H,J]=0$, $C^*_r(A)$ is stable under the Cartan operation of $J$ and we note $C^*_r(A,SO(2))$ the complex of basic cochains of weight $r$. Then we have
\be
C^*(A,SO(2))=\bigoplus_{r=-2}^{\infty}C^*_r(A,SO(2))\ .
\ee
For any cocycle $\om\in C^*_r(A,SO(2))$, 
\be
L_H\om=di_H\om=-r\om
\ee
so that $C^*_r(A,SO(2))$ is acyclic whenever $r\neq 0$. Hence $H^*(A,SO(2))$ is equal to the cohomology of the finite-dimensional subcomplex $C_0^*(A,SO(2))$. The direct computation gives
\be
\begin{array}{lcl}
H^0(A,SO(2)) = \cc &\mbox{with representative}& 1 \\
H^2(A,SO(2)) = \cc &'' & \om^{-1}\om^1\\
H^3(A,SO(2)) = \cc &''&(\om^{-1}\om^1-\omb^{-1}\omb^1)(\om^0+\omb^0)\\
H^5(A,SO(2)) = \cc &''&\om^1\om^{-1}\omb^1\omb^{-1}(\om^0+\omb^0)
\end{array}
\ee
The other cohomology groups vanish.\\

Next we construct a map $C$ from $C^*(A)$ to the bicomplex $(C^{n,m},d_1,d_2)_{n,m\in\zz}$ of \cite{C2} chap. III.2.$\delta$. Let $\Om^m(G_1)$ be the space $m$-forms on $G_1$. $C^{n,m}$ is the space of totally antisymmetric maps $\gamma: {G_2}^{n+1}\rightarrow\Om^m(G_1)$ such that
\be
\gamma(g_0g,...,g_ng)=\gamma(g_0,...,g_n)\circ g\qquad g_i\in G_2,g\in G\ ,\label{inv}
\ee
where $g_ig$ is given by the right action of $G$ on $G_2$, and $G$ acts on $\Om^*(G_1)$ by pullback (left action of $G$ on $G_1$).\\
The first differential $d_1:C^{n,m}\rightarrow C^{n+1,m}$ is
\be
(d_1\gamma)(g_0,...,g_{n+1})=(-)^m\sum_{i=0}^{n+1}(-)^i\gamma(g_0,...,\stackrel{\vee}{g_i},...,g_{n+1})\ ,
\ee
and $d_2:C^{n,m}\rightarrow C^{n,m+1}$ is just the de Rham coboundary on $\Om^*(G_1)$:
\be
(d_2 \gamma)(g_0,...,g_n)= d(\gamma(g_o,...,g_n))\ .
\ee
Of course ${d_1}^2={d_2}^2=d_1d_2+d_2d_1=0$. Remark that for $\gamma\in C^{n,m}$, the invariance property (\ref{inv}) implies
\be
\gamma(g_0,...,g_n)\circ k= \gamma(g_0\triangleleft k,...,g_n\triangleleft k)\qquad \forall k\in G_1\ ,
\ee
in other words the value of $\gamma(g_0,...,g_n)\in\Om^m(G_1)$ at $k$ is deduced from its value at $1$.\\

Let us describe now the construction of $C$. As a vector space, the Lie algebra $A$ is just the direct sum $\mbox{\bf G}_1\oplus{\mbox{\bf G}_2}$, ${\mbox{\bf G}_i}$ being the (complexified) Lie algebra of $G_i$. The cochain complex $C^*(A)$ is then the exterior product $\Lambda A^*=\Lambda {\mbox{\bf G}_1}^*\otimes\Lambda{\mbox{\bf G}_2}^*$. One identifies ${\mbox{\bf G}_1}^*$ with the cotangent space $T^*_1(G_1)$ of $G_1$ at the identity. Since $G_2$ fixes $1\in G_1$, there is a right action of $G_2$ on $\Lambda{\mbox{\bf G}_1}^*$ by pullback. The basis $\{\om^{-1}, \om^0,\omb^{-1},\omb^0\}$ of ${\mbox{\bf G}_1}^*$ is represented by left-invariant one-forms on $G_1$ through the identification
\beq
\om^{-1}&\rightarrow& -\om_{-1}=-y^{-1}dz\qquad \omb^{-1}\ \rightarrow\ -\omb_{-1}=-\yb^{-1}d\zb\non\\
\om^0&\rightarrow& -\om^0=-y^{-1}dy \qquad \omb^0\ \rightarrow\  -\omb^0=-\yb^{-1}d\yb \ ,
\eeq
and the right action of $\psi\in G_2$ reads (cf. (\ref{pull}))
\be
\om^{-1}\cdot \psi=\om^{-1}\ ,\qquad \om^0\cdot\psi =\om^0+\delta_1(\psi)\om^{-1}\ .
\ee
Next, we view a cochain $\om\in C^*(A)$ as a cochain of the Lie algebra of $G_2$ with coefficients in the right $G_2$-module $\Lambda{\mbox{\bf G}_1}^*$. It is represented by a $\Lambda{\mbox{\bf G}_1}^*$-valued right-invariant form $\mu$ on $G_2$. Then $C(\om)\in C^{*,*}$ evaluated on $(g_0,...,g_n)\in{G_2}^{n+1}$ is a differential form on $G_1$ whose value at $1\in G_1$ is
\be
C(\om)(g_0,...,g_n)=\int_{\Delta(g_0,...,g_n)}\mu\quad \in\Lambda T^*_1(G_1)\ ,
\ee
where $\Delta(g_0,...,g_n)$ is the affine simplex in the coordinates $\delta_i,\overline{\delta_i}$, with vertices $(g_0,...,g_n)$. Let $\{\rho_j\}$ be a basis of left-invariant forms on $G_1$. Then 
\be
C(\om)(g_0,...,g_n)=\sum_j p_j(g_0,...,g_n)\rho_j\qquad \mbox{at } 1\in G_1\ ,
\ee
where $p_j(g_0,...,g_n)$ are polynomials in the coordinates $\delta_i,\overline{\delta_i}$. The invariance property (\ref{inv}) enables us to compute the value of $C(\om)(g_0,...,g_n)$ at any $k\in G_1$,
\be
C(\om)(g_0,...,g_n)(k)=\sum_j p_j(g_0\triangleleft k,...,g_n\triangleleft k)\rho_j
\ee
because $\rho_j\circ k=\rho_j$.\\

Connes and Moscovici showed in \cite{CM98} that $C$ is a morphism from $C^*(A,d)$ to the bicomplex $(C^{n,m},d_1,d_2)_{n,m\in\zz}$. In the relative case, it restricts to a morphism from $C^*(A,SO(2),d)$ to the subcomplex $(C^{n,m}_{bas.},d_1,d_2)$ of antisymmetric cochains on $G_2$ with values in the {\it basic} de Rham cohomology $\Om^*(P)=\Om^*(G_1/SO(2))$.\\

It remains to compute the image of $H^*(A,SO(2))$ by $C$. We restrict ourselves to even cocycles, i.e. the unit $1\in H^0(A,SO(2))$ and the first Chern class $c_1\in H^2(A,SO(2))$, defined as the class 
\be
c_1=[2\om^{-1}\om^1]\ .
\ee
One has $C(1)\in C_{bas.}^{0,0}$. The immediate result is
\be
C(1)(g_0)=1\ ,\qquad g_0\in G_2\ .
\ee
For the first Chern class, we must transform $c_1$ into a right-invariant form on $G_2$ with values in $\Lambda T^*_1(G_1)$. We already know that $\om^{-1}$ is represented by $-\om_{-1}=-y^{-1}dz$, which satisfies $\om_{-1}\circ\psi= \om_{-1}$, $\forall\psi\in G_2$. Next, the Taylor expansion of an element $\psi\in G_2$ can be expressed in the coordinates $\delta_n$ thanks to the obvious formula
\be
\ln \psi'(x)= \sum_{n=1}^{\infty} {1\over {n!}}\delta_n(\psi)x^n\ ,\qquad \forall x\in\cc\ .
\ee
One finds:
\be
\psi(x)=x+{1\over 2}\delta_1(\psi)x^2+{1\over {3!}}(\delta_2(\psi)+\delta_1(\psi)^2)x^3+O(x^4)\ .
\ee
It shows that the cochain $\om^1\in C^*(A)$ is represented by the right-invariant 1-form ${1\over 2}d\delta_1$ on $G_2$. Thus at $1\in G_1$, $C(c_1)\in C^{1,1}_{bas.}$ is given by
\beq
C(c_1)(g_0,g_1)&=&\int_{\Delta(g_0,g_1)}-\om_{-1}d\delta_1\non\\
 &=& -\om_{-1}(\delta_1(g_1)-\delta_1(g_0))\qquad g_i\in G_2\ ,
\eeq
and at $k\in G_1$, the 1-form $C(c_1)(g_0,g_1)$ is
\be
C(c_1)(g_0,g_1) =-\om_{-1}(\delta_1(g_1\triangleleft k)-\delta_1(g_0\triangleleft k))\ .
\ee
Since $\om_{-1}=y^{-1}dz$ and $\delta_1(g\triangleleft k)=y\d_z\ln g'(z)$, $z$ and $y$ being the coordinates of $k$, one has explicitly
\be
C(c_1)(g_0,g_1)=-dz(\d_z\ln {g_1}'(z)-\d_z\ln{g_0}'(z))\ .
\ee
It is a basic form on $G_1$ relative to $SO(2)$, then descends to a form on $P=G_1/SO(2)$ as expected.\\

The last step is to use the map $\Phi$ of \cite{C2} theorem 14 p.220 from $(C^{n,m},d_1,d_2)$ to the $(b,B)$ bicomplex of the discrete crossed product $\cinfc(P)\rtimes G_2$. Define the algebra
\be
\Bc=\Om^*(P)\hat{\otimes}\Lambda\cc(G_2')\ ,
\ee
where $\Lambda\cc(G_2')$ is the exterior algebra generated by the elements $\delta_{\psi}$, $\psi\in G_2$, with $\delta_e=0$ for the identity $e$ of $G_2$. With the de Rham coboundary $d$ of $\Om^*(P)$, $\Bc$ is a differential algebra. Now form the crossed product $\Bc\rtimes G_2$, with multiplication rules
\beq
U^*_{\psi}\al U_{\psi} &=& \al\circ\psi\ ,\qquad \al \in\Om^*(P),\psi\in G_2\non\\
U^*_{\psi_1}\delta_{\psi_2}U_{\psi_1}&=& \delta_{\psi_2\circ\psi_1}-\delta_{\psi_1}\ ,\qquad \psi_i\in G_2\ .
\eeq
Endow $\Bc\rtimes G_2$ with the differential $\tilde{d}$ acting on an element $bU^*_{\psi}$ as
\be
\tilde{d}(bU^*_{\psi})=dbU^*_{\psi}-(-)^{\d b}b\delta_{\psi}U^*_{\psi}\ ,
\ee
where $db$ comes from the de Rham coboundary of $\Om^*(P)$. The map
\be
\Phi:(C^{*,*},d_1,d_2)\rightarrow (\cinfc(P)\rtimes G_2, b,B)
\ee
is constructed as follows. Let $\gamma\in C^{n,m}_{bas.}$. It yields a linear form $\tilde{\gamma}$ on $\Bc\rtimes G_2$:
\beq
\tilde{\gamma}(\al\otimes\delta_{g_1}...\delta_{g_n})&=& \int_P \al\wedge\gamma(1,g_1,...,g_n)\ ,\qquad \al\in\Om^*(P),g_i\in G_2\non\\
\tilde{\gamma}(bU^*_{\psi})&=&0\qquad\mbox{if }\ \psi\neq 1\ .
\eeq
Then $\Phi(\gamma)$ is the following $l$-cochain on $\cinfc(P)\rtimes G_2$, $l=\dim P-m+n$
\beq
\Phi(\gamma)(x_0,...,x_l)&=& {{n!}\over{(l+1)!}}\sum_{j=0}^l (-)^{j(l-j)}\tilde{\gamma}(\tilde{d}x_{j+1}...\tilde{d}x_l x_0\tilde{d}x_1...\tilde{d}x_j)\ ,\non\\
&&  x_i\in \cinfc(P)\rtimes G_2\subset \Bc\rtimes G_2\ .
\eeq
The essential tool is that $\Phi$ is a morphism of bicomplexes:
\be
\Phi(d_1\gamma)=b\Phi(\gamma)\ ,\qquad \Phi(d_2\gamma)=B\Phi(\gamma)\ .
\ee
Moreover, if $d_1\gamma=d_2\gamma=0$, $\Phi(\gamma)$ is a cyclic cocycle. This happens in our case. Since $P$ is a $3$-dimensional manifold, the image of $C(1)$ under $\Phi$ is the cyclic 3-cocycle
\be
\Phi(C(1))(x_0,...,x_3)=\int_P x_0dx_1...dx_3\ ,\qquad x_i\in \cinfc(P)\rtimes G_2\ ,
\ee
where $d(fU^*_{\psi})=dfU^*_{\psi}$ for $f\in \cinfc(P)$, $\psi\in G_2$, and the integration is extended over $\Om^*(P)\rtimes G_2$ by setting
\be
\int_P \al U^*_{\psi}=0\qquad \mbox{if}\ \psi\neq 1,\ \al\in\Om^*(P)\ .
\ee

The image of $\gamma=C(c_1)$ is more complicated to compute. One has
\be
\tilde{\gamma}(\al\otimes\delta_g)=-\int_P\al\wedge y^{-1}dz\delta_1(g\triangleleft k)\ ,\qquad\al\in\Om^2(P),g\in G_2
\ee
where $y^{-1}dz\delta_1(g\triangleleft k)=dz\d_z\ln g'(z)$ is, of course, a 1-form on $P$. $\Phi(\gamma)$ is the cyclic 3-cocycle
\beq
\Phi(\gamma)(f_0U^*_{\psi_0},...,f_3U^*_{\psi_3}) &=& -\tilde{\gamma}(f_0U^*_{\psi_0}df_1U^*_{\psi_1}df_2U^*_{\psi_2}f_3\delta_{\psi_3}U^*_{\psi_3} \non\\
&& + f_0U^*_{\psi_0}df_1U^*_{\psi_1}f_2\delta_{\psi_2}U^*_{\psi_2}df_3U^*_{\psi_3}\non\\
&& + f_0U^*_{\psi_0}f_1\delta_{\psi_1}U^*_{\psi_1}df_2U^*_{\psi_2}df_3U^*_{\psi_3})\non
\eeq
\beq
&=& \tilde{\gamma}(f_0\,(df_1\circ\psi_0)\,(df_2\circ \psi_1\psi_0)\,(f_3\circ\psi_2\psi_1\psi_0)\delta_{\psi_2\psi_1\psi_0}\\
&& +f_0\,(df_1\circ\psi_0)\,( f_2\circ\psi_1\psi_0)\, (df_3\circ\psi_2\psi_1\psi_0)(\delta_{\psi_2\psi_1\psi_0}-\delta_{\psi_1\psi_0})\non\\
&& - f_0\, (f_1\circ\psi_0)\, (df_2\circ\psi_1\psi_0)\, (df_3\circ\psi_2\psi_1\psi_0)(\delta_{\psi_1\psi_0}-\delta_{\psi_0}))\ ,\non
\eeq
upon assuming that $\psi_3\psi_2\psi_1\psi_0=\Id$. Using the relation
\be
\delta_1(\psi\triangleleft k)=[\delta_1(U^*_{\psi})U_{\psi}](k)\ ,\qquad\forall k\in G_1,\psi\in G_2
\ee
the computation gives
\be
\Phi(\gamma)(x_0,...,x_3)= \int_P x_0(dx_1dx_2\delta_1(x_3) +dx_1\delta_1(x_2)dx_3 +\delta_1(x_1)dx_2dx_3)y^{-1}dz\ .
\ee

Now recall that $P$ has an invariant volume form $dv_1=e^{2r}dzd\zb dr$. The differential $df$ of a function on $P$ makes use of the horizontal $X=y\d_z$, $\overline{X}=\yb\d_{\zb}$ and vertical $Y+\overline{Y}=-\d_r$ vector fields:
\be
df=y^{-1}dz X.f+\yb^{-1}d\zb \overline{X}.f-dr(Y+\overline{Y}).f\ .
\ee
Then using the relations (\ref{pull}) one sees that $\Phi(C(c_1))$ is a sum of terms involving the Hopf algebra
\be
\Phi(C(c_1))(x_0,...,x_3)= \sum_i\int_P x_0h^i_1(x_1)...h^i_3(x_3)dv_1\ ,
\ee
where the sum $\sum_i h^i_1\otimes h^i_2\otimes h^i_3$ is a cyclic 3-cocycle of $\Hc$ relative to $SO(2)$. This follows from the existence of the characteristic map (\ref{bug})
\be
HC^*(\Hc,SO(2))\rightarrow HC^*(\cinfc(P)\rtimes G_2)
\ee
and the duality between $\Hc$ and $\Hc_*=\cinfc(G_1)\rtimes G_2$ (cf. \cite{CM98}).\\

Returning to the initial situation, where $F$ is the frame bundle of a flat Riemann surface $\Si$, and $P=F/SO(2)$ the bundle of metrics, the above computation shows that the cyclic 3-cocycle on $\Ac_1=\cinfc(P)\rtimes \Gamma$
\be
{[c_1]}(a_0,...,a_3)=\sum_i\int_P a_0h^i_1(a_1)...h^i_3(a_3) dv_1\ ,\qquad a_i\in\Ac_1\ ,
\ee
is the image of $C(c_1)$ by the characteristic map $HC^*(\Hc,SO(2))\rightarrow HC^*(\Ac_1)$. Also the fundamental class
\be
{[P]}(a_0,...,a_3)=\int_P a_0da_1da_2da_3
\ee
is in the range of the characteristic map.\\

Since Connes and Moscovici showed that the Gelfand-Fuchs cohomology $H^*(A,SO(2))$ is isomorphic to the periodic cyclic cohomology of $\Hc$, we have completely determined the odd part of the range of the characteristic map. We can summarize the result in the following \\

\begin{proposition}
Under the characteristic map
\be
H^*(A,SO(2))\simeq H^*(\Hc,SO(2))\rightarrow H^*(\Ac_1)\ ,
\ee
the unit $1\in H^0(A,SO(2))$ maps to the fundamental class $[P]$ represented by the cyclic 3-cocycle
\be
{[P]}(a_0,...,a_3)=\int_P a_0da_1da_2da_3\ ,\qquad a_i\in \Ac_1\ ,
\ee
and the first Chern class $c_1\in H^2(A,SO(2))$ gives the cocycle $[c_1]\in HC^3(\Ac_1)$:
\be
{[c_1]}(a_0,...,a_3)= \int_P a_0(da_1da_2\delta_1(a_3) + da_1\delta_1(a_2)da_3 + \delta_1(a_1)da_2da_3)y^{-1}dz\ .
\ee
\end{proposition}
%\hfill $\Box$\\

In section II we considered an odd $K$-cycle on $C_0(P\times \rr^2)\rtimes\Gamma$ represented by a differential operator $Q'$, which is equivalent, up to Bott periodicity, to an odd $K$-cycle on $C_0(P)\rtimes \Gamma$. $Q'$ is a matrix-valued polynomial in the vector fields $X,\overline{X},Y+\overline{Y}$ and the partial derivatives along the two directions of $\rr^2$. Its Chern character is the cup product
\be
\ch_*(Q')=\varphi\# [\rr^2]
\ee
of a cyclic cocycle $\varphi\in HC^{odd}(\cinfc(P)\rtimes \Gamma)$ by the fundamental class of $\rr^2$. The index theorem of Connes and Moscovici states that $\varphi$ is in the range of the characteristic map (we have to assume that the action of $\Gamma$ on $\Si$ has no fixed point). Hence it is a linear combination of the characteristic classes $[P]$ and $[c_1]$. We shall determine the coefficients by using the classical Riemann-Roch theorem.\\
\vskip 1cm

\noindent {\bf V. A Riemann-Roch theorem for crossed products}\\

We shall first use the Thom isomorphism in $K$-theory \cite{C0}
\be
K_i(C_0(\Si)\rtimes \Gamma)\rightarrow K_{i+1}(C_0(P)\rtimes\Gamma)
\ee
to descend the characteristic classes $[P]$ and $[c_1]$ down to the cyclic cohomology of $\cinfc(\Si)\rtimes \Gamma$. Recall that $C_0(P)\rtimes\Gamma$ is just the crossed product of $C_0(\Si)\rtimes\Gamma$ by the modular automorphism group $\si$ of the associated von Neumann algebra
\be
C_0(P)\rtimes\Gamma = (C_0(\Si)\rtimes\Gamma)\rtimes_{\si}\rr\ .
\ee
By homotopy we can deform $\si$ continuously into the trivial action. For $\la\in [0,1]$, let $\si^{\la}_t=\si_{\la t}$, $\forall t\in\rr$. Then $\si^1=\si$, $\si^0=\Id$ and
\be
(C_0(\Si)\rtimes\Gamma)\rtimes_{\mbox{\small Id}}\rr = C_0(\Si)\rtimes\Gamma \otimes C_0(\rr)\ .
\ee
Next, the coordinate system $(z,\zb)$ of $\Si$ gives a smooth volume form ${{dz\wedge d\zb}\over{2i}}$ together with a representative of $\si$, whose action on the subalgebra $\cinfc(\Si)\rtimes\Gamma$ is
\be
\si_t(fU^*_{\psi})=f|\psi'|^{2it}U^*_{\psi}\ ,\qquad f\in \cinfc(\Si), \psi\in\Gamma\ ,
\ee
and accordingly
\be
\si^{\la}_t(fU^*_{\psi})=f|\psi'|^{2i\la t}U^*_{\psi}\ .
\ee
Remark that the algebra $(C_0(\Si)\rtimes\Gamma)\rtimes_{\si^{\la}}\rr$ is equal to the crossed product $C_0(P)\rtimes_{\la}\Gamma$ obtained from the following deformed action of $\Gamma$ on $P$:
\beq
z&\rightarrow&\psi(z)\qquad \zb\ \rightarrow \ \overline{\psi(z)} \\
r&\rightarrow& r-{1\over 2}\la\ln|\psi'(z)|^2\qquad \psi\in\Gamma\ .\non
\eeq
Hence for any $\la\in [0,1]$, one has a Thom isomorphism
\be
\Phi^{\la}: K_0(C_0(\Si)\rtimes\Gamma)\rightarrow K_1(C_0(P)\rtimes_{\la}\Gamma)\ ,
\ee
and $\Phi^0$ is just the connecting map $K_0(C_0(\Si)\rtimes\Gamma)\rightarrow K_1(S(C_0(\Si)\rtimes\Gamma))$. We introduce also the family $\{[P]^{\la}\}_{\la\in [0,1]}$ of cyclic cocycles
\be
{[P]}^{\la}(a_0^{\la},...,a_3^{\la})=\int_P a_0^{\la}da_1^{\la}...da_3^{\la}\ ,\qquad \forall a_i^{\la}\in \cinfc(P)\rtimes_{\la}\Gamma\ .
\ee
One has $[P]^1=[P]$ and $[P]^0=[\Si]\# [\rr]\in (\cinfc(\Si)\rtimes\Gamma)\otimes \cinfc(\rr)$, where
\be
{[\Si]}(a_0,a_1,a_2)=\int_{\Si} a_0da_1da_2\qquad \forall a_i\in \cinfc(\Si)\rtimes\Gamma \ .
\ee
Moreover for any element $[e]\in K_0(C_0(\Si)\rtimes\Gamma)$ such that $\Phi^{\la}([e])$ is in the domain of definition of $[P]^{\la}$, the pairing
\be
\langle \Phi^{\la}([e]), [P]^{\la}\rangle
\ee
depends continuously upon $\la$. Next for any $\la\in (0,1]$, consider the vertical diffeomorphism of $P$ whose action on the coordinates $(z,\zb,r)$ reads
\be
\tilde{\la}(z)=z\qquad \tilde{\la}(\zb)=\zb\qquad \tilde{\la}(r)=\la r\ .
\ee
Thus for $\la\neq 0$ one has an algebra isomorphism
\be
\chi_{\la}: \cinfc(P)\rtimes_{\la}\Gamma\rightarrow \cinfc(P)\rtimes\Gamma
\ee
by setting
\be 
\chi_{\la}(fU^*_{\psi})= f\circ\tilde{\la}\ U^*_{\psi}\ \qquad \forall f\in\cinfc(P), \psi\in\Gamma\ .
\ee
For any $\la\neq 0$,
\beq
(\chi_{\la})_* \circ \Phi^{\la}&=& \Phi^1\ ,\label{a}\\
(\chi_{\la})^* [P]^1 &=& [P]^{\la}\ .\label{b}
\eeq
Eq.(\ref{a}) comes from the unicity of the Thom map (cf. \cite{C0}), and (\ref{b}) is obvious. Thus $\langle \Phi^{\la}([e]), [P]^{\la}\rangle$ is constant for $\la\neq 0$, and by continuity at $0$
\be
\langle \Phi^1([e]), [P]\rangle = \langle [e], [\Si]\rangle\ .
\ee
This shows that the image of $[P]$ by Thom isomorphism is the cyclic 2-cocycle $[\Si]$ corresponding to the fundamental class of $\Si$. In exactly the same way we show that the image of $[c_1]$ is the cyclic 2-cocycle $\tau$ defined, for $a_i=f_iU^*_{\psi_i}\in\cinfc(\Si)\rtimes\Gamma$, by
\be
\tau(a_0,a_1,a_2)=\int_{\Si} a_0(da_1\d\ln\psi'_2\, a_2+ \d\ln\psi'_1\, a_1da_2)\ ,
\ee
with $\d=dz\d_z$. Note that in the decomposition of the differential on $\Si$, $d=\d+\db$, both $\d$ and $\db$ commute with the pullbacks by the conformal transformations $\psi\in\Gamma$.\\

So far we have considered a {\it flat} Riemann surface and the constructions we made were relative to a coordinate system $(z,\zb)$. We shall now remove this unpleasant feature by using the Morita equivalence \cite{CM98}. In order to understand the general situation, let us first treat the particular case of the Riemann sphere $S^2=\cc\cup\{\infty\}$. We consider an open covering of the sphere by two planes: $S^2=U_1\cup U_2$, $U_1=\cc$, $U_2=\cc$, together with the glueing function $g$:
\beq
g: U_1\backslash\{0\}&\rightarrow& U_2\backslash\{0\}\non\\
z&\mapsto& {1\over z}\ .
\eeq
The pseudogroup of conformal transformations $\Gamma_0$ generated by $\{U^*_g,U_g\}$ acts on the disjoint union $\Si=U_1\amalg U_2$, which is flat. Then $S^2$ is described by the groupoid $\Si\rtimes\Gamma_0$. If $\Gamma$ is a pseudogroup of local transformations of $S^2$, there exists a pseudogroup $\Gamma'$ containing $\Gamma_0$, acting on $\Si$ and such that the crossed product $\cinf(S^2)\rtimes\Gamma$ is Morita equivalent to $\cinfc(\Si)\rtimes\Gamma'$. The latter splits into four parts: it is the direct sum, for $i,j=1,2$, of elements of the form $f_{ij}U^*_{\psi_{ij}}$ with
\be
\psi_{ij}:U_i\rightarrow U_j\quad \mbox{and}\quad \mbox{supp}f_{ij}\subset\mbox{Dom}\psi_{ij}\ .
\ee
For convenience, we adopt a matricial notation for any generic element $b\in\cinfc(\Si)\rtimes\Gamma'$:
\be
b=\left( \begin{array}{cc}
    b_{11} & b_{12} \\
   b_{21} & b_{22} \end{array}\right)\ , \qquad b_{ij}=f_{ij}U^*_{\psi_{ij}}\ .
\ee
Now the Morita equivalence is explicitly realized through the following idempotent $e\in \cinfc(\Si)\rtimes\Gamma'$:
\be
e=\left( \begin{array}{cc}
         {\rho_1}^2 & \rho_1\rho_2 U^*_g \\
    U_g \rho_2\rho_1 & U_g {\rho_2}^2U^*_g 
  \end{array} \right)\ ,\qquad e^2=e\ ,
\ee
where $\{\rho_i\}_{i=1,2}$ is a partition of unity relative to the covering $\{U_i\}$:
\be
\rho_1\in \cinfc(U_1)\ ,\quad {\rho_1}^2+{\rho_2}^2=1\ \mbox{on}\ S^2=U_1\cup \{\infty\}\ .
\ee
The reduction of $\cinfc(\Si)\rtimes\Gamma'$ by $e$ is the subalgebra
\be
(\cinfc(\Si)\rtimes\Gamma')_e = \{b\in \cinfc(\Si)\rtimes\Gamma'/b=be=eb\}\ .
\ee
Its elements are of the form
\be
ebe=\left(\begin{array}{cc}
\rho_1c\rho_1 & \rho_1c\rho_2U^*_g \\
U_g\rho_2c\rho_1 & U_g\rho_2c\rho_2U^*_g
\end{array} \right)
\ee
with $c=\rho_1b_{11}\rho_1 + \rho_2U^*_gb_{21}\rho_1 + \rho_1b_{12}U_g\rho_2 + \rho_2U^*_gb_{22}U_g\rho_2$. Then $c$ can be considered as an element of $\cinf(S^2)\rtimes\Gamma$ under the identification $S^2=U_1\cup\{\infty\}$. $(\cinfc(\Si)\rtimes\Gamma')_e$ and $\cinf(S^2)\rtimes\Gamma$ are isomorphic through the map
\beq
\theta: \cinf(S^2)\rtimes\Gamma&\longrightarrow& (\cinfc(\Si)\rtimes\Gamma')_e \non\\
a &\longmapsto& \left( \begin{array}{cc}
                   \rho_1a\rho_1 & \rho_1a\rho_2U^*_g \\
                   U_g\rho_2a\rho_1 & U_g\rho_2a\rho_2U^*_g \end{array} \right) \ .
\eeq

We are ready to compute the pullbacks of $[\Si]$ and $\tau\in HC^2(\cinfc(\Si)\rtimes\Gamma')$ by $\theta$. This yields the following cyclic 2-cocycles on $\cinf(S^2)\rtimes\Gamma$:
\beq
\theta^*[\Si] &=& [S^2]\ ,\non\\
(\theta^*\tau)(a_0,a_1,a_2)&=& \int_{S^2} a_0\biggl(da_1(\d\ln\psi_2'\,a_2 + [a_2,{\rho_2}^2\d\ln g'])\non\\
&& + \ (\d\ln\psi_1'a_1+ [a_1,{\rho_2}^2\d\ln g'])\biggr)\label{ex}\\
&&-\int_{S^2} a_2a_0a_1 d({\rho_2}^2)\d\ln g'\ ,\non
\eeq
with $a_i=f_iU^*_{\psi_i}\in \cinf(S^2)\rtimes\Gamma$. In formula (\ref{ex}), $S^2=U_1\cup\{\infty\}$ is gifted with the coordinate shart $(z,\zb)$ of $U_1$, which makes sense to $\psi_i'(z)=\d_z\psi_i(z)$ and $g'(z)=\d_zg(z)=-1/z^2$, but gives singular expressions at $0$ and $\infty$. We can overcome this difficulty by introducing a smooth volume form $\nu=\rho(z,\zb){{dz\wedge d\zb}\over{2i}}$ on $S^2$. The associated modular automorphism group $\si^{\nu}$ leaves $\cinf(S^2)\rtimes\Gamma$ globally invariant and is expressed in the coordinates $(z,\zb)$ by
\be
\si^{\nu}_t(fU^*_{\psi}) =\left({{\nu\circ\psi}\over{\nu}}\right)^{it}fU^*_{\psi}= \big({{\rho\circ\psi}\over{\rho}}|\d_z\psi|^2\big)^{it}fU^*_{\psi}\ ,\qquad \forall t\in\rr\ .
\ee
Define the derivation $\delta^{\nu}$ on $\cinf(S^2)\rtimes\Gamma$
\beq
\delta^{\nu}(fU^*_\psi)&\equiv& -i[\d,{d\over{dt}}\si^{\nu}_t](fU^*_{\psi})\vert_{t=0}\non\\
&=& [\d,\ln\big({{\rho\circ\psi}\over{\rho}}|\d_z\psi|^2\big)](fU^*_{\psi})\\
&=& \d\ln\psi'\, fU^*_{\psi} - [\d\ln\rho, fU^*_{\psi}]\ .
\eeq
One has
\be
\d\ln\psi'\, fU^*_{\psi}+[fU^*_{\psi},{\rho_2}^2\d\ln g'] = \delta^{\nu}(fU^*_{\psi})+ [\d\ln\rho - {\rho_2}^2\d\ln g',fU^*_{\psi}]\ ,
\ee
where the 1-form $\om=\d\ln\rho - {\rho_2}^2\d\ln g'$ is globally defined, nowhere singular on $S^2$. Let $R^{\nu}=\d\db\ln\rho$ be the curvature 2-form associated to the K\"ahler metric $\rho dz\otimes d\zb$. One has the commutation rule
\be
(\db\delta^{\nu}+\delta^{\nu}\db)a = [R^{\nu},a]\qquad \forall a\in \cinf(S^2)\rtimes\Gamma\ .
\ee
Simple algebraic manipulations show that the following 2-cochain
\be
\tau^{\nu}(a_0,a_1,a_2)=\int_{S^2}a_0(da_1\delta^{\nu}a_2+\delta^{\nu}a_1da_2) + \int_{S^2}a_2a_0a_1R^{\nu}\label{eu}
\ee
is a cyclic cocycle. Moreover, $\tau^{\nu}$ is cohomologous to $\theta^*\tau$. To see this, let $\varphi$ be the cyclic 1-cochain
\be
\varphi(a_0,a_1)=\int_{S^2} (a_0da_1-a_1da_0)\om\ .
\ee
Then for all $a_i\in \cinf(S^2)\rtimes\Gamma$,
\beq
(\tau^{\nu}-\theta^*\tau)(a_0,a_1,a_2)&=& -\int_{S^2}(a_0da_1a_2+a_2da_0a_1+ a_1da_2a_0)\om\non\\
&=& b\varphi(a_0,a_1,a_2) \ .
\eeq

It is clear now that the construction of characteristic classes for an arbitrary (non flat) Riemann surface $\Si$ follows exactly the same steps as in the above example. Using an open cover with partition of unity, one gets the desired cyclic cocycles by pullback. Choose a smooth measure $\nu$ on $\Si$, then the associated modular group is
\be
\si^{\nu}_t(fU^*_{\psi})=\left({{\nu\circ\psi}\over{\nu}}\right)^{it}fU^*_{\psi}\qquad fU^*_{\psi}\in\cinfc(\Si)\rtimes\Gamma\ .
\ee
The corresponding derivation
\be
D^{\nu}(fU^*_{\psi})=\ln\left({{\nu\circ\psi}\over{\nu}}\right)fU^*_{\psi}
\ee
allows to define the noncommutative differential
\be
\delta^{\nu}=[\d,D^{\nu}]\ .
\ee
Then the characteristic classes of the groupoid $\Si\rtimes\Gamma$ are given by $[\Si]$ and $[\tau^{\nu}]\in HC^2(\cinfc(\Si)\rtimes\Gamma)$, where $\tau^{\nu}$ is given by eq.(\ref{eu}) with $S^2$ replaced by $\Si$. \\

In the case $\Gamma=\Id$, the crossed product reduces to the commutative algebra $\cinfc(\Si)$ for which $(\delta^{\nu}=0)$
\be
\tau^{\nu}(a_0,a_1,a_2)= \int_{\Si}a_0a_1a_2R^{\nu}
\ee
is just the image of the cyclic 0-cocycle
\be
\tau_0^{\nu}(a)=\int_{\Si}aR^{\nu}
\ee
by the suspension map in cyclic cohomology $S:HC^*(\cinfc(\Si))\rightarrow HC^{*+2}(\cinfc(\Si))$. Thus the periodic cyclic cohomology class of $\tau^{\nu}$ corresponds in de Rham homology to the cap product
\be
{1\over{2\pi i}}[\tau^{\nu}]=c_1(\kappa)\cap [\Si] \quad\in H_0(\Si)
\ee
of the first Chern class of the holomorphic tangent bundle $\kappa$ by the fundamental class. This motivates the following definition:\\

\begin{definition}
Let $\Si$ be a Riemann surface without boundary and $\Gamma$ a discrete pseudogroup acting on $\Si$ by local conformal transformations. Let $\nu$ be a smooth volume form on $\Si$, and $\si^{\nu}$ the associated modular automorphism group leaving $\cinfc(\Si)\rtimes\Gamma$ globally invariant. Then the Euler class $e(\Si\rtimes\Gamma)$ is the class of the following cyclic 2-cocycle on $\cinfc(\Si)\rtimes\Gamma$
\be
{1\over{2\pi i}}\tau^{\nu}(a_0,a_1,a_2)={1\over{2\pi i}}\int_{\Si}(a_2a_0a_1R^{\nu}+a_0(da_1\delta^{\nu}a_2+\delta^{\nu}a_1da_2))\ ,
\ee
where $\delta^{\nu}$ is the derivation $-i[\d,{d\over{dt}}\si^{\nu}_t\vert_{t=0}]$, and $R^{\nu}$ is the curvature of the K\"ahler metric determined by $\nu$ and the complex structure of $\Si$. Moreover, this cohomology class is independent of $\nu$.
\end{definition}

Now if $\Gamma=\Id$, the operator $Q$ of section II defines an element of the $K$-homology of $\Si\times\rr^2$. It corresponds to the tensor product of the classical Dolbeault complex $[\db]$ of $\Si$ by the signature complex $[\si]$ of the fiber $\rr^2$, so that its Chern character in de Rham homology is the cup product
\beq
\ch_*(Q) &=& \ch_*([\db])\# \ch_*([\si]) \non\\
&=& ([\Si]+{1\over 2}c_1(\kappa)\cap [\Si])\# 2[\rr^2]\quad \in H_*(\Si\times\rr^2)
\eeq
which yields, by Thom isomorphism, the homology class on $\Si$
\be
2[\Si] + c_1(\kappa)\cap [\Si]\qquad \in H_*(\Si)\ .
\ee
Next for any $\Gamma$, we know from the last section that the Chern character of the Dolbeault $K$-cycle, expressed in the periodic cyclic cohomology of $\cinfc(\Si)\rtimes\Gamma$, is a linear combination of $[\Si]$ and $e(\Si\rtimes\Gamma)$. Thus we deduce immediately the following generalisation of the Riemann-Roch theorem:\\

\begin{theorem}
Let $\Si$ be a Riemann surface without boundary and $\Gamma$ a discrete pseudogroup acting on $\Si$ by local conformal mappings without fixed point. The Chern character of the Dolbeault $K$-cycle is represented by the following cyclic 2-cocycle on $\cinfc(\Si)\rtimes\Gamma$
\be
\mbox{\normalsize ch}_*(Q)=2[\Si]+e(\Si\rtimes\Gamma)\ .
\ee
\hfill $\Box$
\end{theorem}

\noindent {\bf Acknowledgements:} I am very indebted to Henri Moscovici for having corrected an erroneous factor in the final formula.

%\makeatletter
%\def\@biblabel#1{#1.\hfill}

\end{document}